\documentclass[10pt,aps,prd,twocolumn,floats,floatfix,showpacs,superscriptaddress,nofootinbib]{revtex4-2}
\usepackage[utf8]{inputenc} 
\usepackage{graphicx,mathtools,amssymb,amsmath,amsthm,amsfonts,epsfig,epsf,bm,,multirow}
\usepackage[outdir=./]{epstopdf}
\usepackage[usenames]{color}
\usepackage{tensor}
\usepackage{csquotes}
\usepackage{mathrsfs}
\usepackage{autobreak}
\usepackage{physics}
\usepackage{cancel}
\usepackage[font=small,labelfont=bf, justification=Justified,format=plain]{caption}
\usepackage{slashed}
\usepackage{pifont}
\usepackage[dvipsnames]{xcolor}
\usepackage[colorlinks=true]{hyperref}
\usepackage{cleveref}

\def\del{\partial}

\def\be{\begin{equation}}
\def\ee{\end{equation}}
\def\bea{\begin{eqnarray}}
\def\eea{\end{eqnarray}}

\usepackage[normalem]{ulem}

\begin{document}

\title{Higher co-dimension de Sitter branes}

\author{\textbf{Florian Niedermann}}
\email{florian.niedermann@su.se}
\affiliation{Nordita, KTH Royal Institute of Technology and Stockholm University\\
Hannes Alfv\'ens v\"ag 12, SE-106 91 Stockholm, Sweden}

\author{\textbf{Antonio Padilla}}
\email{antonio.padilla@nottingham.ac.uk}
\affiliation{Nottingham Centre of Gravity, University of Nottingham,
University Park, Nottingham NG7 2RD, United Kingdom}
\affiliation{School of Physics and Astronomy, University of Nottingham, University Park, Nottingham NG7 2RD, United Kingdom}


\begin{abstract}
We extend the arguments of Maldacena and N\'u\~nez to include higher co-dimension brane setups and derive a new no-go theorem. Specifically, we show that under reasonable assumptions on the energy-momentum conservation and the bulk curvature, co-dimension-two branes fail to support stable de Sitter solutions. For co-dimensions higher than two embedded in a compact internal space, we show that negative tension sources would be required. This result places strong constraints on the viability of higher-dimensional braneworld models as a means to obtain de Sitter space within string theory.

\end{abstract}

\maketitle

\section{Introduction}
A fundamental goal of string theory is to provide a UV-complete framework for cosmology. Given the observed accelerated cosmic expansion at both early and late times \cite{Planck:2018vyg, SupernovaCosmologyProject:1998vns, SupernovaSearchTeam:1998fmf}, an obvious goal is to find a natural embedding of de Sitter space. However, finding stable de Sitter solutions within string compactifications has proven notoriously difficult. A variety of no-go theorems, beginning with the classical results of Maldacena and N\'u\~nez \cite{Maldacena:2000mw}, demonstrate that under broad assumptions such as the presence of a smooth compactification with controlled internal curvature, string theory struggles to accommodate de Sitter vacua. These results have been further reinforced by more recent swampland conjectures, including the de Sitter conjecture \cite{Obied:2018sgi, Garg:2018reu, Ooguri:2018wrx}, which suggests that any consistent low-energy effective theory emerging from string theory should either avoid dS space altogether or include instabilities leading to rapid decay. Of course, recent results from the DESI collaboration \cite{DESI:2025zgx} point towards a dynamical form for dark energy \cite{Copeland:2006wr}. However, it is, perhaps, even more difficult to construct robust quintessence mechanisms within string theory \cite{Garg:2018zdg,Cicoli:2018kdo,Hebecker:2019csg,ValeixoBento:2020ujr, Cicoli:2021fsd,Cicoli:2021skd,Hebecker:2023qke}. See \cite{Cicoli:2024yqh} for an explicit construction of inflation at early times and quintessence at late,  albeit with a brane uplift. 

Efforts to construct explicit de Sitter vacua have generally relied on introducing ingredients such as flux compactification and warped throats,  with perturbative and non-perturbative corrections \cite{Kachru:2003aw, Balasubramanian:2005zx}. However, these scenarios have been subject to growing scrutiny, with concerns about their consistency within a fully UV-complete setup \cite{Banks:2003es, Danielsson:2018ztv}. Given these challenges, alternative mechanisms for realizing de Sitter-like spacetimes in string theory have gained traction. This includes the so-called $\alpha'$ complete cosmologies, which attempt to go beyond perturbative string theory \cite{Hohm:2019ccp,Hohm:2019jgu, Bernardo:2019bkz,Bernardo:2020zlc,Bernardo:2019bkz, Nunez:2020hxx,Liu:2024sgb}. However, it is clear that they suffer from generic instabilities, as might have been expected from world-sheet no-go theorems \cite{Kutasov:2015eba}.  

Another interesting direction includes braneworld scenarios where our four-dimensional universe arises as a dynamical brane embedded in a higher-dimensional bulk.  Model building with co-dimension-one branes has enjoyed something of a renaissance of late especially in view of de Sitter phenomenology, most notably through the dark bubble and/or end-of-the-world branes~\cite{Muntz:2024joq,Banerjee:2018qey,Banerjee:2019fzz,Banerjee:2020wix,Banerjee:2020wov}. However, here our interest lies in higher co-dimension brane set-ups, with two or more co-dimensions. Although these are much more difficult to study,  the interplay between bulk curvature and localized energy-momentum sources can often lead to novel effects, such as the relaxation of the four-dimensional vacuum energy via self-tuning mechanisms \cite{Navarro:2003vw,Carroll:2003db,Burgess:2011mt,Burgess:2011va,Niedermann:2014bqa,Bernardo:2022cck}. Recent work has explored whether co-dimension-two set-ups can naturally lead to de Sitter solutions \cite{Burgess:2024jkx}.

In this paper, we shall extend the Maldacena-N\'u\~nez argument to braneworld setups with higher co-dimension branes. By analyzing the structure of the Einstein equations in the presence of such branes, we derive a new no-go theorem that rules out de Sitter solutions for co-dimension-two branes under very mild assumptions.  For higher co-dimensions beyond two, we find the need for negative effective brane tensions. Our  work builds upon earlier extensions of the Maldacena-N\'u\~nez  theorem \cite{Steinhardt:2008nk} and highlights the remaining obstacles in the search for a consistent string-theoretic realization of a de Sitter universe.

The paper is organized as follows. In Section \ref{sec:nogo}, we extend the Maldacena-Nu\~nez no-go theorem to general higher co-dimension brane setups, deriving a constraint equation for the on-brane curvature under broad assumptions. We then evaluate the near-brane contributions in the limit of vanishing regulator, showing that they either vanish or diverge, thereby ruling out de Sitter solutions for co-dimension-two branes and requiring negative tension sources for higher co-dimensions. In Section \ref{sec:co-dim-2}, we apply our analysis to a well-studied six-dimensional supergravity model recently claimed to support de Sitter branes \cite{Burgess:2024jkx}. We demonstrate how our no-go result reproduces and strengthens earlier findings, including the vanishing of the on-brane curvature in the thin-brane limit.  We conclude in Section \ref{sec:conc} with a discussion of implications, assumptions, and possible loopholes—particularly those related to finite thickness effects.

\section{A no go theorem} \label{sec:nogo}

We consider a $D=d+n$ dimensional spacetime with coordinates $X^A=(x^\mu, y^a)$, where $x^\mu$ are the $d$-dimensional coordinates for the external space and $y^a$ describe the $n$-dimensional internal space. The metric ansatz is:
\begin{equation}\label{eq:metric}
    ds^2 = e^{2\phi(y 
    )} \bar{\gamma}_{\mu\nu}(x) dx^\mu dx^\nu + \hat g_{ab}(y)dy^a dy^b \,,
\end{equation}
where $\bar \gamma_{\mu\nu}$ is assumed to be the metric for a maximally symmetric spacetime. The Ricci tensor components are explicitly computed as:
\begin{align}
    R^\mu_\nu &= e^{-2\phi} \bar{R}^\mu_\nu - \delta^\mu_\nu \left( \hat \nabla^2 \phi+d (\hat \nabla \phi)^2  \right), \label{Rmn} \\
    R^a_b &= \hat  R^a_b -d \left(\hat \nabla^a \hat \nabla_b \phi+\hat \nabla^a \phi \hat\nabla_b \phi \right).
\end{align}
From the Einstein equations, we have that
\begin{equation}
    R^A_B=T^A_B-\frac{1}{D-2}T\delta^A_B\,,
\end{equation}
from which we obtain the constraint $R^\mu_\mu=\tau$ where 
\begin{equation}
    \tau=\frac{1}{D-2} \left[(n-2)T^\mu_\mu-d\, T^a_a \right]\,.
\end{equation}
Using \eqref{Rmn}, this yields 
\begin{equation}\label{eq:Einstein}
    e^{-2\phi} \bar{R} - e^{-d \phi} \hat \nabla^2 e^{d\phi}=\tau\,.
\end{equation}
We now multiply through by $e^{d \phi}$ and integrate over a general section of the internal space, $\mathcal{M}$, implying
\begin{equation}
    \bar R J(\mathcal{M})-\int_{\partial \mathcal{M}} dS_{n-1} n_a \hat \nabla^a e^{d \phi}=K(\mathcal{M}) \label{KM}\,,
\end{equation}
where
\begin{eqnarray}
    J(\mathcal{M})&=&\int_\mathcal{M} dV_n e^{(d-2)\phi}\,,  \\
    K(\mathcal{M})&=&\int_\mathcal{M} dV_n e^{d\phi}\tau \label{K}\,.
\end{eqnarray}
Here $dV_n=\sqrt{\hat g} d^n y$ is the volume measure on the internal space section, $\mathcal{M}$,  and $dS_{n-1}$ the measure on its boundary $\partial\mathcal{M}$.

We now separate the full internal space, $\mathcal{M}_\text{tot}$  into a compact region $\mathcal{M}_\text{out}$ containing no brane sources and  compact regions $\mathcal{M}_i$ containing the $i$th brane source.  It follows that 
\begin{equation}
    \mathcal{M}_{\text{tot}} = \mathcal{M}_{\text{out}} + \sum_i \mathcal{M}_i \,.
\end{equation}
Of course,  the full internal space is also compact, and we further assume that it has no overall boundary, $\partial  \mathcal{M}_{\text{tot}} =0$.  We then see that 
\begin{equation}
   \partial \mathcal{M}_{\text{out}} =-\sum_i \partial \mathcal{M}_i\,.
\end{equation}
Since 
\begin{eqnarray}
    \bar R J(\mathcal{M}_\text{out})-\int_{\partial \mathcal{M}_\text{out}} dS_{n-1} n_a \hat \nabla^a e^{d \phi} &=& K(\mathcal{M}_\text{out})\,,\,\,\, \\\
    \bar R J(\mathcal{M}_i)-\int_{\partial \mathcal{M}_i} dS_{n-1} n_a \hat \nabla^a e^{d \phi}&=&K(\mathcal{M}_i)\,,
\end{eqnarray}
we infer that
\begin{equation}
    \bar R\left[J(\mathcal{M}_\text{out})+\sum_i J(\mathcal{M}_i)\right]=K(\mathcal{M}_\text{out})+\sum_i K(\mathcal{M}_i) \label{bar R}\,,
\end{equation}
or equivalently
\begin{equation}\label{eq:Rbar}
    \bar RJ(\mathcal{M}_\text{tot})=K(\mathcal{M}_\text{out})+\sum_i K(\mathcal{M}_i)\,.
\end{equation}
Since $\mathcal{M}_\text{tot}$ is compact, it is clear that $J(\mathcal{M}_\text{tot})>0$ and finite.\footnote{{We stress that models with infinite bulk volume are not covered by our arguments.}} Like Maldacena and N\'u\~nez \cite{Maldacena:2000mw}, we  also assume $\tau \leq 0$, albeit away from the brane sources, giving $K(\mathcal{M}_\text{out}) \leq 0$. In the absence of the brane contributions given by $K(\mathcal{M}_i)$, we immediately recover the Maldacena-N\'u\~nez theorem, which yields $\bar R \leq 0$, ruling out de Sitter solutions on the external space. Here, we extend the theorem by including contributions from regions close to the brane sources.  In these regions we no longer make any assumptions  about the sign of $\tau$. However, we do assume conservation of the energy–momentum tensor. As we will discuss shortly, the energy–momentum tensor in these near-brane neighbourhoods incorporates the effects of both bulk fields and regularised brane sources. We will later show that these near-brane contributions either diverge or vanish in the limit where any regulator is removed.

To explore the near-brane contributions more closely, we choose coordinates in a small neighbourhood of the $i$th brane so that metric on the internal space is
\begin{equation}\label{metric_internal}
    \hat g_{ab}(y)dy^a dy^b=dr^2 + A^2(r) q_{ij}(\theta) d\theta^i d\theta^j\,.
\end{equation}
Here the brane is assumed to lie at $r=0$, with $r$ measuring the radial distance from the source. $A(r)$ measures the size of the compact $n-1$ dimensional space and vanishes as $r \to 0$ in agreement with having a higher co-dimensional defect. The compact space is typically a sphere with angular coordinates $\theta^i$  and metric $q_{ij}(\theta)$. For a sufficiently small neighbourhood, we further assume that $\phi=\phi(r)$. 

The value of $\phi$ at the brane is unphysical and can be scaled to any value we choose. This is because a shift $\phi \to \phi+c$ for some (possibly divergent) constant $c$ can be absorbed by a constant rescaling of the metric $\bar \gamma_{\mu\nu} \to e^{-2c}\bar \gamma_{\mu\nu} $. Note that the physical brane curvature, $e^{-2\phi(0)}\bar R$, is invariant under these shifts. Furthermore, since $\bar R \to e^{2c} \bar R$, $J \to e^{(d-2)c}J$ and $K \to e^{dc} K$, it follows that both sides of the curvature equation \eqref{bar R} pick up an overall factor of $e^{dc}$, which can be cancelled. We will use this freedom to assume, without loss of generality, that $\phi(r)$ remains finite as $r\to 0^+$.

What about the energy momentum tensor? As we alluded to earlier, in the neighbouhood of the brane, there could be contributions from bulk fields as well as localised contributions from the brane itself.   For higher co-dimension branes, the latter must be regularised in some way. For example, for a co-dimension-two brane, we may consider a co-dimension-one brane of fixed radius and then take the radius to be very small~\cite{Peloso:2006cq,Burgess:2008yx,Niedermann:2018kki,Niedermann:2015via,Niedermann:2015vbk}.  More generally, we assume that the (regularised) energy momentum tensor has the following non-vanishing components
\begin{equation}
    T^A_B=(T^\mu_\nu(r), T^r_r(r), T^i_j(r))
\end{equation}
in the neighbourhood of the brane, depending only on the radial coordinate. By integrating the regularised source over this small neighbourhood, we obtain  the effective energy-momentum of the source, 
\begin{equation}
    T_\text{eff}{}^A_B=\int_{\mathcal{M}_i} d^n y \sqrt{\hat g} T^A_B\,,
\end{equation}
which must be finite for a successful regularisation. Henceforth we take $\mathcal{M}_i$ to be the region defined by $0 \leq r \leq\epsilon$, for some arbitrarily small $\epsilon>0$.  It follows that
\begin{equation}\label{eq:Teff}
    T_\text{eff}{}^A_B=\Omega_{n-1} \int_0^\epsilon dr A^{n-1} T^A_B
\end{equation}
where $\Omega_{n-1}=\int d^{n-1}\theta \sqrt{q}$ is the volume of the compact angular directions. 

We now want to estimate the near-brane contribution to our formula \eqref{bar R} for the brane curvature. To this end,  we exploit the energy-conservation equation 
\begin{equation}
    \partial_r T^r_r + \frac{A'}{A} \left[(n-1)T^r_r - T^i_i \right] + \phi' (dT^r_r - T^\mu_\mu) = 0.
\end{equation}
This allows us to eliminate the angular contribution $T^i_i$ in our expression for $K(\mathcal{M}_i)$. In particular, we find that 
\begin{multline}
     \tau =
\frac{1}{D-2} \left[\left(n-2+\frac{d\phi'A}{A'}\right)T^\mu_\mu \right. \\ 
\left.
-\frac{dA}{A'} \del_r T^r_r-d\left(n+\frac{d\phi'A}{A'}\right) T^r_r 
\right]\,.
\end{multline}
Plugging this into our expression \eqref{K} for $K(\mathcal{M}_i)$, we obtain 
\begin{multline}\label{eq:K}
  K(\mathcal{M}_i) = \frac{e^{d \phi(0)}}{D-2} \left[(n-2)T_\text{eff}{}^\mu_\mu \right.\\
  \left. -\Omega_{n-1}(\Delta_1(\mathcal{M}_i)+\Delta_2(\mathcal{M}_i))
  \right]\,,
\end{multline}
where  we define
\begin{eqnarray}
    \Delta_1(\mathcal{M}_i)&=&\int_0^\epsilon dr A^{n-1} \omega(r) T^\mu_\mu \,,\\
     \Delta_2(\mathcal{M}_i)&=&d e^{-d \phi(0)}
   \int_0^\epsilon dr \frac{\del_r (A^n e^{d\phi} T^r_r)}{A'}\,, 
\end{eqnarray}
and a weighting function
\begin{multline}
    \omega(r)=(n-2)\left[1-e^{d(\phi(r)-\phi(0))}\right]\\-e^{d(\phi(r)- \phi(0))} d \phi' \frac{A}{A'}\,.
\end{multline}
Near the regularised brane, we take $A \approx A^{(\epsilon)}_0 r^\alpha$ and $\phi \approx \phi^{(\epsilon)}_0+\phi^{(\epsilon)}_1 r^\beta$, where $\alpha, \beta >0$ and the constant coefficients depending on the regularisation scale. $\phi^{(\epsilon)}_0$ is assumed to remain finite and non-vanishing as we remove the regulator, $\epsilon \to 0^+$.  As explained earlier, this assumption is carried out without loss of generality thanks to the freedom to shift, $\phi \to \phi +c$, for any constant $c$. We impose even weaker constraints on $\phi^{(\epsilon)}_1$ and $A^{(\epsilon)}_0 > 0$, requiring that $A^{(\epsilon)}_0 \epsilon^{\alpha - 1}$ remains finite and that $\phi^{(\epsilon)}_1 \epsilon^\beta \to 0$ as $\epsilon \to 0^+$. The first condition ensures that the gradient $A'(\epsilon)$ remains finite in the limit $\epsilon \to 0^+$ {(in accordance with having a genuine co-dimension-$n$ defect)}, while the second guarantees that $\phi(\epsilon) \to \phi_0$.

We now estimate the weighting function, 
\begin{equation}
    \omega(r) \approx -\left(n-2+\frac{\beta}{\alpha}\right) d\, \phi_1^{(\epsilon)} r^\beta
\end{equation}
which is vanishingly small in the limit that $\epsilon \to 0^+$ for $0\leq r\leq \epsilon$. Although $T^\mu_\mu$ can and does diverge in this limit, it does so while guaranteeing that $T_\text{eff}{}^\mu_\mu=\Omega_{n-1} \int_0^\epsilon dr A^{n-1} T^\mu_\mu$ remains finite.  The presence of a vanishingly small weighting function in $\Delta_1$ will generically\footnote{This will certainly be true if the regulated source $T^\mu_\mu$ doesn't flip sign for $0\leq r \leq \epsilon$, as can be easily proved using the triangle inequality.} force the integral to vanish as we remove the regulator.  We conclude that $\Delta_1 \to 0$ as $\epsilon \to 0^+$. 

The second integral, $\Delta_2$, requires a little more work. To estimate the contribution of $T^r_r$, we make use of the Einstein equation,
\begin{multline}\label{rr-Einstein}
    T^r_r=\frac12 \left[
    (n-1)(n-2) \left(\frac{A'}{A}\right)^2 -\frac{R(q)}{A^2}-\bar R e^{-2\phi} 
    \right. \\ \left.
    +2(n-1) d \phi' \frac{A'}{A} +d (d-1) \phi'{}^2\right]\,,
\end{multline}
where $R(q)>0$ is the Ricci curvature of the compact space with metric $q_{ij}$. It follows that in a neighbourhood of the brane, 
\begin{eqnarray}
    T^r_r &\approx& \sum_i \lambda_i r^{-2\gamma_i} \nonumber\\
    &\approx& \frac12 \left[ (n-1)(n-2)\alpha^2 r^{-2} -\frac{R(q)}{\left(A_0^{(\epsilon)}\right)^2}r^{-2\alpha}-\bar Re^{-2\phi_0} \right.\nonumber\\
    &&  +2(n-1) d \alpha \beta \phi_1 r^{\beta-2}+d(d-1)\beta^2 \phi_1^2 r^{2\beta-2} \Bigg]\,.
  \end{eqnarray}
This yields 
\begin{equation}
    \Delta_2 (\mathcal{M}_i) \approx \left( A^{(\epsilon)}_0\right)^{n-1} d \sum_i\lambda_i s_i \left[ \frac{ r^{\rho_i}}{\rho_i} \right]^\epsilon_0 \label{Delta}
\end{equation}
with $s_i=n-2\gamma_i/\alpha$ and $\rho_i=\alpha(s_i-1)+1$. Note that this result has a divergence at the origin if and only if $\rho_i \leq 0$ and $\lambda_i s_i \neq 0$   for any $i$.  If this is {\it not} the case, the contribution at the origin vanishes, and we are left with
\begin{equation}
    \Delta_2 (\mathcal{M}_i) \approx \left( A^{(\epsilon)}_0\right)^{n-1} d \sum_i\lambda_i s_i  \frac{ \epsilon^{\rho_i}}{\rho_i} =\frac{d W(\epsilon) }{2\alpha^n}   \epsilon^{n-2}\,, \label{Delta2}
\end{equation}
where 
\begin{eqnarray}
    W(\epsilon) &=& \left(A'(\epsilon)\right)^{n-1}  \left[ \frac{(n-1)(n-2)\alpha^2 (\alpha \, n-2)}{\alpha(n-1)-1} \right.\nonumber
    \\
    &&-\frac{\alpha \, n}{\alpha(n-1)+1} \bar Re^{-2\phi_0} \epsilon^2\nonumber\\
    &&  + \frac{2 \, d\,(n-1) \, \alpha \,\beta \,(\alpha\, n+\beta-2)}{\alpha(n-1)  +\beta-1}\left[\phi(\epsilon)-\phi_0\right] \nonumber \\
   && +d(d-1)\frac{\beta^2 (\alpha \, n+2\beta-2)}{\alpha(n-1)+2\beta-1} \left[\phi(\epsilon)-\phi_0\right]^2 \Bigg] \nonumber \\
  &&  -\left(A'(\epsilon)\right)^{n-3} R(q) \frac{(n-2)\alpha^3}{\alpha(n-3)+1}\,.
\end{eqnarray}
As we remove the regulator, $\epsilon \to 0^+$, recall that  we required that $\phi(\epsilon) \to \phi_0$ and that $A'(\epsilon)$ remains finite.  For $n=2$, we see that the first and last terms vanish identically in our expression for $W(\epsilon)$, with the remaining terms vanishing in the limit $\epsilon \to 0^+$. Meanwhile, for $n \geq 3$, $W(\epsilon)$ remains finite.  It follows that for $n \geq 2$, the right hand side of \eqref{Delta2} vanishes as $\epsilon \to 0^+$. 

We therefore have two possibilities: as we remove the regulator, either $\Delta_2$ diverges or it vanishes.  In fact, this outcome might have been anticipated: $\Delta_2$ can  be understood as capturing brane contributions from internal directions transverse to the brane—such as radial and angular directions. In the regulated system, these contributions can be finite and non-zero. However, as the regulator is removed, symmetry considerations suggest that such off-brane contributions must either vanish or diverge—and indeed, they do.

If there is a divergence in $\Delta_2$, there is also a divergence in the brane curvature, via equation \eqref{bar R}, so we dismiss this as unphysical. In contrast, if there is no divergence, we have  $\Delta_1, \Delta_2 \to  0$ as $\epsilon \to 0^+$ and so 
\begin{equation}\label{K_final}
  K(\mathcal{M}_i) \approx \frac{e^{d \phi_0}}{D-2} (n-2)T_\text{eff}{}^\mu_\mu\,.
\end{equation}
For higher co-dimensions ($n \geq 3$),  the brane sources give a positive contribution to the brane curvature if and only if $T_\text{eff}{}^\mu_\mu >0$. For effective brane tensions $T_\text{eff}{}^\mu_\nu=-\sigma \delta^\mu_\nu $, the tension must be negative $\sigma<0$. Actually, negative tension objects exist in string theory in the form of O-planes \cite{Dai:1989ua}, although they correspond to fixed points of an orientifold symmetry and are not dynamical objects. More speculatively, negative tension branes have also been proposed as formal, supersymmetric objects that arise via T-duality on a timelike circle \cite{Hull:1998vg,Dijkgraaf:2016lym}. For co-dimension-two branes ($n=2$), we see that \eqref{K_final} vanishes and fails to give the positive contribution required to support a positive curvature on the brane. In this case, we might hope that subleading contributions come in with the right sign to support de Sitter solutions. However, it is clear that these contributions are vanishingly small in the limit where $\epsilon \to 0^+$.

This completes our no-go theorem, ruling out the possibility of de Sitter vacua on co-dimension-two branes. For branes of higher co-dimension, the presence of negative tension objects appears to be necessary. An obvious loophole, however, is to retain the regulator and allow for branes of finite thickness. Strictly speaking, such objects are no longer truly of co-dimension-two or higher. To see how finite-thickness branes might give rise to de Sitter solutions, let us return to equation~\eqref{Delta2}, and assume \( \rho_i > 0 \) and \( \lambda_i s_i \neq 0 \), while keeping \( \epsilon \) finite. It is straightforward to see that when \( \lambda_i s_i < 0 \), \( \Delta_2 \) acquires a finite negative value. Through equation~\eqref{bar R}, this contributes positively to the brane curvature, thereby opening up the possibility of de Sitter solutions on branes of finite thickness.

\section{A closer look at co-dimension-two models} 
\label{sec:co-dim-2}
To show how all of this plays out in a concrete setting, we look at the EFT of a six-dimensional model that has been recently claimed to exhibit a de Sitter vacuum on its co-dimension-two branes~\cite{Burgess:2024jkx}. The model arises as a particular limit of chiral gauged 6D (1,0) supergravity~\cite{Nishino:1986dc,Randjbar-Daemi:1985tdc} and has been studied extensively in the literature~(see for example~\cite{Salam:1984cj,Aghababaie:2003wz,Gibbons:2003di,Tolley:2005nu}). Its action reads\footnote{This formulation is already restricted to solutions where a 3-form field strength associated with the Kalb-Ramond field and the fermionic particle content have been set to zero.}
\begin{multline}\label{eq:action}
S =  \int_\mathcal{M} d^6X \sqrt{-g} \Big\{\frac{1}{2 \kappa^2} \left[R_6 - \partial_A \varphi \partial^A \varphi \right] \\
-\frac{1}{4} e^{-\varphi} F_{AB} F^{AB} - \frac{2 e^2}{\kappa^4} e^\varphi \Big\} + S_\mathrm{branes}[\varphi] \,,
\end{multline}
where $\varphi$ is a dilaton and $F_{AB} =2 \partial_{[A}A_{B]}$ is the field strength of a 1-form gauge potential with components $A_B$. The model features two 3-branes that are placed at the north (``N'') and south pole (``S'') of a compact two-dimensional internal space $\mathcal{M}_\mathrm{tot}$ with worldvolume action
\begin{align}
S_\mathrm{brane}[\varphi] = \sum_{b \in \left\{\mathrm{N}, \mathrm{S} \right\}}  \int d^4x \sqrt{-\hat g}\, \sigma_b(\varphi)\,,
\end{align}
where we introduced the $\varphi$-dependent tension $\sigma_b(\varphi)$. This brane-dilaton coupling is a distinctive feature of the model, which leads to a soft breaking of classical scale invariance that vanishes in the limit where $\sigma_b^\prime(\varphi) \to 0$. 

For solving the coupled bulk-brane system, we look for configurations with $O(2)$ symmetry in the internal space and maximal symmetry along the brane directions. The metric of the internal space in \eqref{metric_internal} thus reads
\begin{equation}
    \hat g_{ab}(y)dy^a dy^b=dr^2 + A^2(r)  d\theta^2\,,
\end{equation}
where for concretness we place the south pole brane at $r=0$ and its northern counterpart at $r=\pi L$ with $L$ setting the length scale of the compact space. For the dilaton and Maxwell field, we correspondingly 
set $\varphi(r)$ and 
$A = A_\theta(r) d\theta$.

The six dimensional  Maxwell equation then becomes
\begin{align}
\partial_A \left(\sqrt{-g} e^{-\varphi} F^{AB} \right) = 0 ,
\end{align}
which integrates to $F_{r\theta} = Q  A(r) e^{\varphi(r)-4\phi(r)}$ with $Q$ an integration constant.  This angular flux provides the bulk pressure needed to stabilise the internal space. With this, the dilaton equation of motion reads
\begin{multline}\label{dilaton}
- \frac{1}{\kappa^2} \frac{e^{-4\phi}}{A} \left(e^{4\phi}A \varphi^\prime\right)^\prime \\
= \frac{e^\varphi}{2}\left(Q^2 e^{-8 \phi} - \frac{4 e^2}{\kappa^4} \right)- \sum_b \frac{\delta_b^{(\epsilon)}}{2\pi A} \sigma'_b(\varphi)\,,
\end{multline}
where we introduced a common regularisation that replaces the brane with a ring in extra space of radius $\epsilon$; explicitly $\delta_b^{(\epsilon)}= \delta(r-\epsilon)$~\cite{Peloso:2006cq,Burgess:2008yx,Niedermann:2018kki} (for an alternative approach that resolves the brane as a vortex, see~\cite{Burgess:2015nka,Burgess:2015gba}).
Moreover, Eq.~\eqref{eq:Einstein} evaluates to
\begin{align}\label{eq:barR_6D}
e^{-2\phi}\bar{R}   - \frac{4e^{-4\phi}}{A } \left(A e^{4\phi} \phi^\prime \right)^\prime = -  \left(T_r^r + T_\theta^\theta \right)\,,
\end{align}
where
\begin{align}
T_r^r &=\frac12 \varphi'^2+\frac12 e^\varphi \kappa^2 \left(Q^2 e^{-8 \phi} - \frac{4 e^2}{\kappa^4} \right)\,, \label{Tr}  \\
T_\theta^\theta &=-\frac12 \varphi'^2+ \frac12 e^\varphi \kappa^2 \left(Q^2 e^{-8 \phi} - \frac{4 e^2}{\kappa^4} \right) +  \sum_b \left( T_\mathrm{b}^{(\epsilon)}\right)^\theta_\theta \,, \label{Ttheta}
\end{align}
and so
\begin{align}\label{Tr+Ttheta}
T_r^r + T_\theta^\theta = e^\varphi \kappa^2 \left(Q^2 e^{-8 \phi} - \frac{4 e^2}{\kappa^4} \right) +  \sum_b \left( T_\mathrm{b}^{(\epsilon)}\right)^\theta_\theta \,.
\end{align}
The last term corresponds to a localised angular source component, which naturally arises for branes with a finite transverse width. For the ring regularisation, it takes the simple form
\begin{align}\label{EMT_brane_theta_theta}
\left(T_b^{(\epsilon)}\right)^\theta_\theta =   \kappa^2 \sum_b \frac{\delta_b^{(\epsilon)}}{2\pi A} p^{(\epsilon)}_{b}\,,
\end{align}
where we introduced the angular pressure $p^{(\epsilon)}_{b}$ along the compact brane direction. Using definition \eqref{eq:Teff}, it can be identified with the effective energy-momentum tensor, \textit{i.e.}, $ \kappa^2 p^{(\epsilon)}_b = T_\text{eff}{}^\theta_\theta$. There has been some debate as to whether or not this term can give a finite non-vanishing contribution in the limit $\epsilon \to 0$. While in \cite{Niedermann:2015via,Niedermann:2015vbk} it was found to vanish, the authors in~\cite{Burgess:2015kda} (and recently in~\cite{Burgess:2024jkx}) {seem} to maintain that it can give a finite, non-vanishing contribution. As we will see, our no-go theorem provides an independent argument supporting the former claim.

In any event, integrating \eqref{eq:barR_6D} over the whole internal space $\mathcal{M}_\mathrm{tot}$ yields 
\begin{multline}
\bar{R} J(\mathcal{M}_\mathrm{tot}) = -\int dr A e^{4\phi+\varphi }\kappa^2 \left(Q^2 e^{-8 \phi} - \frac{4 e^2}{\kappa^4} \right)\\
-\sum_b e^{4\phi_b} \kappa^2 p^{(\epsilon)}_b\,,
\end{multline}
where a subscript ``b'' denotes evaluation at the position of the brane, e.g.\  $\phi_b = \phi(r_b)$.  Substituting the dilaton equation and using that the total derivative evaluates to zero when integrated over $\mathcal{M}_\mathrm{tot}$, we obtain
\begin{align}\label{R_6D}
\bar{R} J(\mathcal{M}_\mathrm{tot}) = -\kappa^2  \sum_b e^{4\phi_b} \left(2 \sigma_b^\prime(\varphi_b) +  p^{(\epsilon)}_b  \right)\,.
\end{align}
This result relates the brane curvature $\bar{R}\,e^{-2 \phi_b} $ to the local properties of the branes. As mentioned before, we only need to demand that the physical brane curvature is finite, but we can employ the shift $\phi \to \phi+c$ alongside the rescaling $\bar{\gamma}_{\mu\nu} \to e^{-2c}\,\bar{\gamma}_{\mu\nu} $ to ensure that also $\phi_b$ is finite\footnote{Note that this must be accompanied by a shift $Q \to Qe^{4c}$, so that the field strength and corresponding flux quantisation condition remain invariant. This also preserves the form of  \eqref{R_6D}.}.

To make further progress with this expression, we 
consider the brane matching conditions, relating the radial derivative of the bulk fields to the source properties. 
This is particularly simple in the case of our ring regularisation, which replaces the co-dimension-two brane by a co-dimension-one defect placed at radial position $r=r_b \equiv \epsilon$. The benefit of this approach is that we can impose elementary flatness along the symmetry axis at $r=0$,
\begin{align}\label{eq:bc_axis}
A(0)=0 \,, \quad A'(0)=1\,, \quad \phi'(0)=0\,, \quad \mathrm{and} \quad \varphi'(0)=0\,.
\end{align}
Israel's junction conditions~\cite{Israel:1966rt} then match the inside of the ring to the outside. Their four-dimensional and angular components are
\begin{align}\label{Israel}
\left[K^\mu_\nu \right]_\pm - \delta^\mu_\nu \left[ K_\alpha^\alpha +  K_\theta^\theta\right]_\pm &=   \frac{\kappa^2}{2\pi A_b} \, \sigma_b \, \delta^\mu_\nu\,, \nonumber \\
\left[ K_\alpha^\alpha\right]_\pm &=   \frac{\kappa^2}{2\pi A_b} \, p_\theta^{(\epsilon)} \,,
\end{align}
where we introduced
$
[f]_\pm = f(\epsilon^+)-f(\epsilon^-)
$
to denote the discontinuity of any function $f(r)$ across the brane, and the extrinsic curvature tensor evaluates to
\begin{align}\label{eq:extrinsic_K}
K_\alpha^{\beta} = \phi'\, \delta^\beta_\alpha\,, \quad \text{and} \quad K_\theta^\theta = \frac{A'}{A}\,.
\end{align}
We also recall that $\sigma_b$ and $p_\theta^{(\epsilon)}$, which denote the brane's tension and internal angular pressure, respectively, are related to the effective energy momentum tensor in \eqref{eq:Teff} through $T_\mathrm{eff}{}^\mu_\nu = -\sigma_b\, \delta^\mu_\nu$ and $T_\mathrm{eff}{}^\theta_\theta = p_\theta^{(\epsilon)}$.\footnote{For the delta-function regularization employed here, the integration in \eqref{eq:Teff} is understood as extending to $\epsilon^+$.}

We can now use the regularity of the geometry around the symmetry axis to extend \eqref{eq:bc_axis} up to the inside surface of the brane, e.g.\ $A(\epsilon^-) = A(0) + \mathcal{O}(\epsilon)$. Moreover, only radial derivatives of bulk fields are discontinuous across the brane. From \eqref{Israel} together with the definitions in \eqref{eq:K}, we then obtain (valid up to corrections of order $\epsilon$)
\begin{align}\label{matching_conditions}
4\, A_b \,\phi_b' &= \frac{\kappa^2}{2\pi}\, p_b^{(\epsilon)}\, , \nonumber \\
A_b' &= \left(1-\frac{\kappa^2\, \sigma_b}{2\pi} \right)  - \frac{3}{4} \frac{\kappa^2}{2\pi}\, p_b^{(\epsilon)}\,.
\end{align}
A third matching condition follows directly from integrating the dilaton equation in \eqref{dilaton} over an infinitesimal radial interval around $r=\epsilon$:
\begin{align}
 A_b \varphi_b' &= \frac{\kappa^2}{2\pi} \sigma_b' \,.
\end{align}
Next, we consider the radial Einstein equation
\begin{multline}\label{radial_constraint}
 6 \phi'^2 + 4 \phi' \frac{A'}{A}-\frac{1}{2} \varphi'^2 \\
 = \frac{e^\varphi}{2} \kappa^2 \left(Q^2 e^{-8 \phi} - \frac{4 e^2}{\kappa^4} \right) +\frac{\bar{R}}{2} e^{-2 \phi}\,.
\end{multline}
Since this equation is first-order in derivatives, it can be used to further constraint the on-brane source properties. To that end, we multiply the equation by $A(r)^2$ and evaluate at it at $r=r_b\equiv \epsilon$. As argued before, for $\epsilon \to 0^+$, we can assume without loss of generality that  $A_b^2 e^{-2\phi_b} \to 0$. Moreover, to make contact with the results in the literature, e.g.~\cite{Burgess:2024jkx, Niedermann:2015vbk}, we make the simplifying assumption that  $A_b^2 e^{\varphi_b} \to 0$.\footnote{We stress that our more general argument does not rely on this assumption.} As a consequence the right hand side of \eqref{radial_constraint} vanishes at leading order in $\epsilon$, whereas the left hand side provides us with a non-trivial constraint on the brane-localised source terms. To be specific, substituting \eqref{matching_conditions}, and solving for $p_b^{(\epsilon)}$ yields (for subcritical tensions $\kappa^2 \sigma \leq 2\pi$)
\begin{multline}\label{pressure}
\kappa^2  p^{(\epsilon)}_b  = \frac{4}{3} \Big\{ \left( 2 \pi - \kappa^2 \sigma_b \right)\\
\pm \sqrt{\left( 2 \pi - \kappa^2 \sigma_b \right)^2 - \frac{3}{4} \kappa^4 \left(\sigma_b^\prime \right)^2} \Big\}\,,
\end{multline}
where the negative root is chosen to recover the correct scale invariant limit for $\sigma_b'(\varphi_b) \to 0$.
Both expressions \eqref{R_6D} and \eqref{pressure} agree with previous findings in the literature~\cite{Bayntun:2009im,Niedermann:2015vbk}. In particular, the authors of~\cite{Burgess:2024jkx} exploit them to argue for a de Sitter geometry on the brane {(compare Eqs.~\eqref{R_6D} and \eqref{pressure} to their Eqs.~(3.20) and (3.19), respectively)}. At first sight, this looks indeed promising: if we can arrange for a situation where 
\begin{align}\label{constraint_param}
- 2 \left( 2 \pi - \kappa^2 \sigma_b \right)< \sqrt{3} \sigma_b'\kappa^2 < 0 \,,
\end{align}
the above equations imply a positive brane curvature alongside a positive \textit{non-vanishing} angular pressure $p_b^{(\epsilon)}$.  The problem with this is that $\sigma_b(\varphi_b)$ and $\sigma_b'(\varphi_b)$ depend on the bulk profiles through their dependence on $\varphi_b = \varphi(r_b)$. The bulk profiles are in turn entirely fixed by the requirement that the fields satisfy the correct boundary conditions in \eqref{matching_conditions} at \textit{both} branes. However, these boundary conditions depend themselves on $\sigma_b(\varphi_b)$ and $\sigma_b'(\varphi_b)$. This leads to very implicit constraints on the values these coupling parameters can take, questioning if the de Sitter condition can be satisfied.    

The authors in \cite{Niedermann:2015vbk} accounted for that difficulty by solving the full brane-bulk system numerically for a specific choice of brane-dilaton couplings. As a result, they found that $\bar{R} \to 0$ for $\epsilon \to 0^+$ (corresponding to $p^{(\epsilon)}_b \to 0$). Here, we take another path closer to the arguments presented in the previous section. As before, we take $A(r) = A_0^{(\epsilon)} \,r^\alpha$ and $\phi(r)= \phi_0^{(\epsilon)} + \phi_1^{(\epsilon)} \, r^\beta$, where $\alpha, \beta > 0$. Substituting this into \eqref{matching_conditions} and solving for $A_0^{(\epsilon)}$ and $\phi_1^{(\epsilon)}$, we obtain at $r=\epsilon$
\begin{align}\label{coefficients}
A_0^{(\epsilon)} &= \frac{a_0}{\alpha} \epsilon^{1-\alpha}\,,\nonumber \\
\phi_1^{(\epsilon)} &= \frac{\alpha}{4 \beta} \frac{1}{a_0} \frac{\kappa^2}{2\pi} \epsilon^{-\beta} p_b^{(\epsilon)}\,,
\end{align}
where 
\begin{align}
a_0 = 1- \frac{\kappa^2}{2\pi} \left(\sigma_b + \frac{3}{4}p_b^{(\epsilon)}\right)\,.
\end{align}
Demanding that the brane tension $\sigma_b$ has to remain finite for $\epsilon \to 0^+$, we conclude from  \eqref{pressure} and \eqref{constraint_param} that also the pressure $p_b^{(\epsilon)}$ and hence $a_0$ have to be finite. As a result, $A'(\epsilon)$ is finite too, in agreement with an ordinary radially symmetric geometry and consistent with our assumptions from the previous section. Moreover, the finiteness of $\phi(0)$---which we can demand as argued before without loss of generality---implies that\footnote{We can exclude the case where $\phi_1^{(\epsilon)} \epsilon^\beta \to \mathrm{const} $, as this can be accommodated by a trivial shift of $\phi_0^{(\epsilon)}$. 
} 
$\phi_1^{(\epsilon)} \epsilon^\beta \to 0 $.  We then infer from \eqref{coefficients} that $p_b^{(\epsilon)} \to 0$ as $\epsilon \to 0^+$. This is where we seem to deviate from the analysis in~\cite{Burgess:2024jkx}, which appears to assume a non-vanishing pressure. In any event, for $p_b^{(\epsilon)} = 0$ to be compatible with~\eqref{pressure}, we have to demand $\sigma_b'(\varphi_b) = 0$, which, due to \eqref{R_6D}, implies a vanishing on-brane curvature in accordance with our general argument.

Alternatively, and as argued in the previous section, one could hope to retain a finite brane thickness and thus a finite pressure $p^{(\epsilon)}_b$. However, in that case the on-brane curvature -- an observable IR quantity -- is controlled by the value of the UV regulator $\epsilon$ (see~\cite{Niedermann:2015vbk} for explicit results), which one might argue goes against the spirit of effective field theories.

\section{Conclusions} \label{sec:conc}

In this work, we have extended the Maldacena-Nu\~nez no-go theorem to higher co-dimension brane setups and derived robust constraints on the existence of de Sitter vacua in such models. Our analysis rests on several mild but well-motivated assumptions: a compact internal space, supergravity in the bulk (i.e.\ $\tau \leq 0$), finite on-brane curvature, and the presence of genuine co-dimension-$n$ sources—i.e.\ delta-function-like defects with no finite width. We have also worked entirely within the framework of standard General Relativity.

Under these assumptions, we showed that co-dimension-two branes cannot support de Sitter curvature, while higher co-dimension branes require sources with negative effective tension. These can in principle be realised by orientifold planes in string theory~\cite{Dai:1989ua}, though such objects are non-dynamical. This result places strong constraints on braneworld approaches to de Sitter model building and highlights the difficulty of obtaining positive cosmological constant solutions in higher-dimensional theories with localised sources.

Nevertheless, an intriguing loophole emerges when the assumption of infinitesimal brane thickness is relaxed. Finite-thickness effects can be systematically incorporated via induced gravity terms on the brane, as shown in \cite{Garfinkle:1989mv,Gregory:1989gg}. These corrections modify the junction conditions, introducing new contributions that depend explicitly on the brane curvature. Within our formalism, such effects correspond to curvature depending contributions to $T^\mu_\mu$. It is notable that an induced Einstein Hilbert term on the brane mimics negative tension when the brane curvature is positive. 

However, for higher co-dimenson defects, these induced gravity terms are often accompanied by ghost-like instabilities~\cite{Dubovsky:2002jm,Hassan:2010ys}. That said, the universality of these ghost modes has been disputed, at least for large enough brane tensions \cite{Berkhahn:2012wg,Niedermann:2014bqa,Eglseer:2015xla}. This suggests that finite brane thickness, encoded via induced curvature corrections, presents a promising direction for future investigation. While not guaranteed to circumvent our no-go theorem, this loophole provides a technically well-defined and physically motivated framework to explore whether regulated sources can support metastable de Sitter solutions. A detailed study of such configurations, particularly their stability and embedding in string compactifications, may shed new light on the possibility of realising de Sitter space in higher-dimensional braneworld scenarios.

\acknowledgements
We are grateful to Cliff Burgess, Thomas Van Riet and Ziqi Yan for useful comments and discussions. AP would like to thank Arne Slot for $N=20$.  FN was supported by VR Starting Grant 2022-03160 of the Swedish Research Council and AP by STFC consolidated grant number ST/T000732/1. For the purpose of open access, the authors have applied a CC BY public copyright licence to any Author Accepted Manuscript version arising. No new data were created during this study.

\bibliography{ref}

\end{document}